\documentclass{aa}
\usepackage{graphicx, natbib,epsfig,amsmath,amssymb, mathrsfs, txfonts}
\DeclareGraphicsExtensions{.eps, .ps}
\usepackage{epstopdf}
\bibpunct{(}{)}{;}{a}{}{,}
\begin{document}
\title{Atmospheric image blur  with finite outer
  scale \\ or partial adaptive correction}

\author{P. Martinez\inst{1} \and J. Kolb\inst{1}  \and A. Tokovinin\inst{2} \and
  M. Sarazin\inst{1}}
\institute{European Southern Observatory, Karl-Schwarzschild-Strasse 2, D-85748, Garching, Germany
\and Cerro-Tololo Inter American Observatory, Casilla 603, La Serena, Chile} 
\offprints{P. Martinez}

\abstract
{}
{Seeing-limited resolution in large telescopes working over wide
  wavelength range depends substantially on the turbulence outer scale
  and cannot be adequately described by one ``seeing'' value. We attempt to clarify frequent confusions on this matter.}
{We study  the effects  of finite turbulence  outer scale  and partial
  adaptive  corrections  by   means  of  analytical  calculations  and
  numerical simulations. }
{If a von K\`arm\`an turbulence  model is adopted, a simple approximate
  formula  captures   the  dependence  of   atmospheric  long-exposure
  resolution on the outer scale over the entire practically interesting
  range  of  telescope  diameters  and  wavelengths. In the infrared (IR), the difference
  with the standard Kolmogorov seeing formula can exceed a factor of
  two. We find that low-order adaptive turbulence correction produces
  residual wave-fronts with effectively small outer scale, so even
  very low compensation order leads to a substantial improvement in
  resolution over seeing, compared to the standard theory.} 
{Seeing-limited resolution  of large telescopes,  especially in the
IR,  is currently  under-estimated  by not  accounting  for the  outer
scale.  On  the  other  hand,  adaptive-optics  systems  designed  for
diffraction-limited imaging  in the IR  can improve the resolution  in the
visible by as much as two times.}


\maketitle

\section{Introduction}
\label{sec:intro}

Image blur of astronomical objects caused by terrestrial atmosphere is
traditionally called ``seeing''. In the 2-nd half of the 20-th century
this phenomenon has been understood and quantified \citep{Young74}. This
understanding was based on  considering the distorted wave-fronts as a
random   stationary  process   with  a   power-law  spectrum   --  the
Kolmogorov-Obukhov model  \citep{Tatarskii61, Roddier81}. This  theory describes the  shape of
the  atmospheric long-exposure  Point Spread  Function (PSF)  and many
other phenomena  by a single  parameter, e.g. the  Fried's coherence
radius  $r_0$ \citep{Fried66}.  The  theory  predicts dependence  of  the  PSF
Full-Width  at  Half   maximum  (FWHM)  $\varepsilon_{0}$  on  wavelength
$\lambda$ and $r_0$:
\begin{equation}
\varepsilon_{0} = 0.976 \; \lambda /r_{0} .
\label{eq1}
\end{equation}
In  this paper  we  assume  that $r_0$  and  $\varepsilon_0$ refer  to
observations at zenith.  By  adopting a standard wavelength $\lambda =
500$\,nm, we  can replace $r_0$  with $\varepsilon_0$ and  this single
parameter is nowadays usually called  ``seeing''. Here we use the term
{\em seeing} in this precise sense, meaning $\varepsilon_0$ at 500\,nm
at zenith.

The success of this theory led most people to believe that the atmospheric
parameters $r_0$ or $\varepsilon_0$ actually exist and can be measured
with high accuracy,  given adequate means.  In fact  the match between
real physical quantities like  PSF or various statistical estimates of
distorted  wave-fronts to  the Kolmogorov-Obukhov theory  varies from
very good  to poor,  but it  is never perfect.  The concept  of seeing
becomes questionable  if we push  it too far. 

The  physics of  turbulence implies  that the  spatial  power spectral
density (PSD) of phase distortions  $W_{\phi} ({\bf f})$ (${\bf f}$ is
the spatial frequency in m$^{-1}$) deviates from the pure power law at
low  frequencies.   A popular  von  K\`arm\`an  (vK) turbulence  model
\citep{Tatarskii61, Ziad2000, Conan} introduces an additional
parameter, the {\em outer scale} $L_0$:
\begin{equation}
W_{\phi}  ({\bf f}) = 0.0229\;  r_0^{-5/3} \; ( |{\bf f}|^2 + L_0^{-2}  )^{-11/6} .
\label{eq:L0}
\end{equation}
Equation~\ref{eq:L0} is the definition  of $L_0$.  The Kolmogorov model
corresponds to  $L_0 = \infty$.  In  the vK model,  $r_0$ describes the
high-frequency asymptotic behavior of the spectrum, and thus loses its
sense  of  an  equivalent  wavefront  coherence  diameter  as  defined
originally by  \citet{Fried66}. Obviously, Eq.~\ref{eq1} is  no longer valid
as well. 

It  remains an  open question  whether wave-front  statistics actually
correspond  to Eq.~\ref{eq:L0}.  Proving  the   vK model
experimentally would be a difficult and eventually futile goal because
large-scale     wavefront    perturbations     are     anything    but
stationary. However, it is  firmly established that the phase spectrum
does  deviate from the  power law  \citep{Ziad2000, Toko2007}. The  Eq.~\ref{eq:L0} with
additional parameter  $L_0$ provides a  useful first-order description
of this behavior. Existing  experimental data on $L_0$ are interpreted
here  in this sense.

In this paper, we study the modifications of Eq.~\ref{eq1} implied by
the finite outer scale. Our analytical calculations are confirmed by
extensive numerical simulations. We show that for finite $L_0$ the
atmospheric FWHM becomes smaller than predicted by  Eq.~\ref{eq1}, and
that this difference can be substantial. The practical consequences
for operation of large telescopes are discussed.  
The  lack  of low-frequency  power  is typical  not  only  for the  vK
turbulence,  but also for  partially corrected  wave-fronts resulting,
e.g.,   from  tip-tilt  compensation   (fast  guiding)   or  low-order
adaptive-optics  (AO)  correction.  Such  a correction  leads  to  small
effective $L_0$.  We apply the  same analytical treatment to this case
and study the shrinking of the PSF halo under partial AO
compensation.

\section{Analytical treatment}
\label{sec:an}

The calculation of the long-exposure PSF is done by multiplying the
telescope optical transfer function (OTF) by an additional term, the
{\em atmospheric OTF}:
\begin{equation}
T_a ({\bf u}) =  \exp [ -0.5 D_{\phi}(\lambda {\bf u})], 
\label{eq:Tf}
\end{equation}
where  ${\bf  u}$  is  the  angular spatial frequency  (in  inverse  radians),
$\lambda$ is the  imaging wavelength, and $ D_{\phi}({\bf  r})$ is the
phase structure  function (SF) \citep{Goodman85, Roddier81}.  This expression is
general,  applicable  to any  turbulence  spectrum  and any  telescope
diameter. In the case of a  large ideal telescope with diameter $D \gg
r_0$ the  diffraction can be  neglected and the long-exposure  OTF and
PSF are accurately described by Eq.~\ref{eq:Tf}.

The analytic  expression for the  phase structure function in  the von
K\`arm\`an model can be found in \citep{Conan, Consortini72,Toko2002}. 
For infinite $L_0$, it transforms into  $ D_{\phi}(r) = 6.88 (r/r_0)^{5/3}$. 

\begin{figure}[!t]
\begin{center}
\includegraphics[width=9cm]{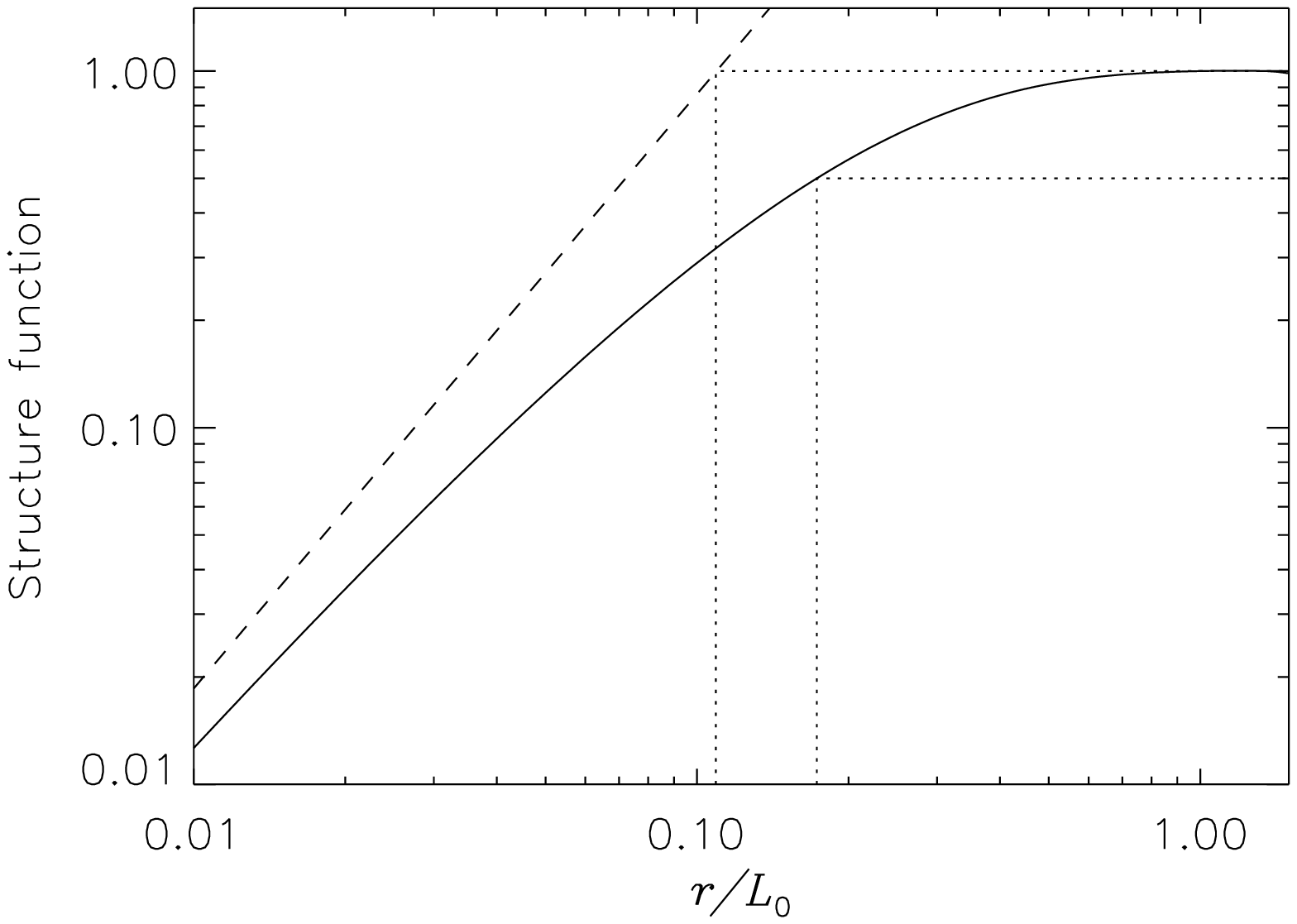}
\includegraphics[width=9cm]{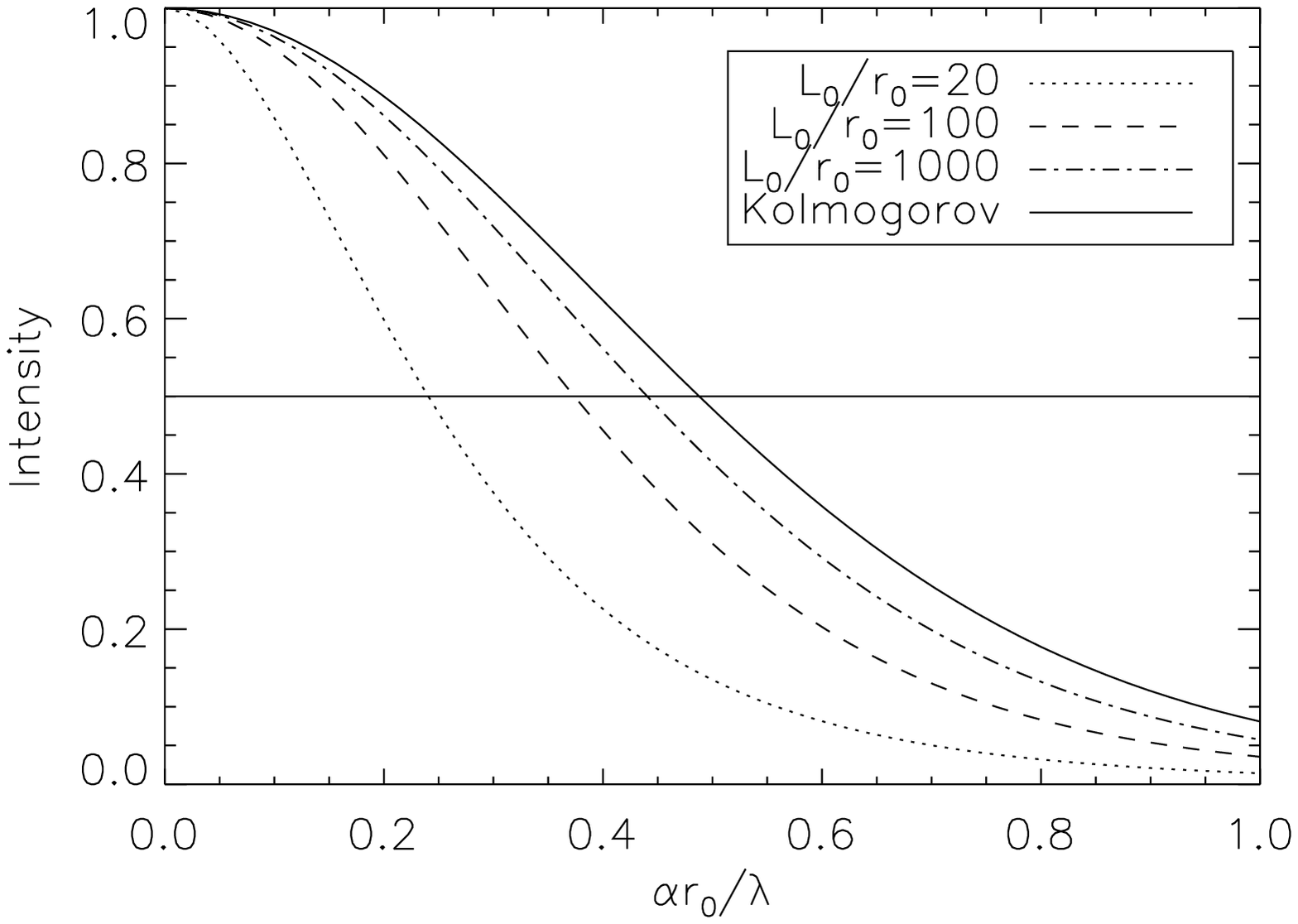}
\end{center}
\caption{Top panel:  comparison of the von K\`arm\`an  (full line) and
  Kolmogorov (dashed line) phase structure functions with the same
  $r_0$. 
Bottom panel: normalized atmospheric PSFs for different $L_0/r_0$ ratios. }
\label{fig:vK}
\end{figure} 

Figure~\ref{fig:vK} (top) plots the SFs for Kolmogorov and vK with the
same  $r_0$.  In  the  latter case  the  SF saturates  at  $r >  L_0$,
reaching asymptotically  the level $0.17  (L_0/r_0)^{5/3}$. It reaches
half-saturation at  $r = 0.17 L_0$.   The Kolmogorov SF  with the same
$r_0$ crosses the  vK saturation level at $r =  0.109 L_0$. This tells
us that the  effect of finite outer scale is  strong at distances much
shorter than $L_0$,  and that it would be  misleading to compare $L_0$
directly with the telescope diameter.

Putting  the vK SF into  Eq.~\ref{eq:Tf},  we find  that for  finite
$L_0$  $T_a  ({\bf  u})$ does  not  go  to  zero at  large  arguments,
therefore its  inverse Fourier transform  (the PSF) formally  does not
exist. However, in the case when $r_0 \ll L_0$ this level is small and
it can be neglected. In  Fig.~\ref{fig:vK} (bottom) we compare the PSF profiles
for different values of $L_0/r_0$, including $L_0 = \infty$.

A first-order  approximation of the  FWHM of atmospheric PSFs ($\varepsilon_{\rm vK}$) under vK turbulence  has
been suggested by  \citet{Toko2002}: 
\begin{equation}
\varepsilon_{\rm vK} \approx \varepsilon_{0} \; \sqrt{1 - 2.183 \; 
  ( r_{0} / L_{0})^{0.356}} .
\label{toko}
\end{equation}
\noindent 
This  formula  is valid  for  $L_0/r_0  > 20  $  to  an accuracy  of
$\pm$1\%. We remind the reader that while $r_0$ depends on the wavelength, $L_0$ does not. 
At smaller $L_0/r_0$  values, the atmospheric PSF develops a
strong  core-halo  structure,  and  its  FWHM becomes  less  and  less
meaningful.  The  actual PSF  in a telescope  is a convolution  of the
atmospheric blur  with diffraction, aberrations,  guiding errors, etc.
As neither of these factors is described by a Gaussian, calculation of
the combined  FWHM as a  quadratic sum of individual  contributions is
not accurate. 

\noindent Similarly a formula for the FWHE, half-energy diameter ($\beta_{vk}$), can be derived with the same accuracy \citep{Toko2002}:
\begin{equation}
\beta_{\rm vK} \approx \beta_{0} \; \sqrt{1 -  1.534\; 
  ( r_{0} / L_{0})^{0.347}} .
\label{toko2}
\end{equation}
\noindent 
where $\beta_0$ is the diameter of the circle that contains one-half of the total PSF energy in the Kolmogorov theory ($\beta_0 = 1.15\lambda/r_0$).

The following section gathers results obtained with extensive numerical simulations to confirm the reliability and validity domain of Eq. \ref{toko}, and thereby Eq. \ref{toko2}. 

\section{Numerical simulations}
\label{assumptions}

\subsection{Random wavefronts}

The atmospheric  turbulence is generated with  1000 uncorrelated phase
screens on $8192  \times 8192$ array equivalent to  a 100 meters width
physical size (pixel size  12.2\,mm).  The principle of the generation
of   a    phase   screen   is   based   on    the   Fourier   approach
\citep{McGlamery87}: randomized  white noise  maps are colored  in the
Fourier space by the turbulence PSD (Eq. \ref{eq:L0}); the inverse Fourier
transform  of an outcome  corresponds to  a phase  screen realization.
The  large  size  of  the  simulated phase  screens  is  mandatory  to
correctly sample the $L_{0}$ and  to compute PSF for large telescopes.
The simulations consider  several $L_{0}$ cases (10, 22,  32.5, 50, 65
m and $\infty$).
\begin{figure*}[!t]
\begin{center}
\includegraphics[width=9cm]{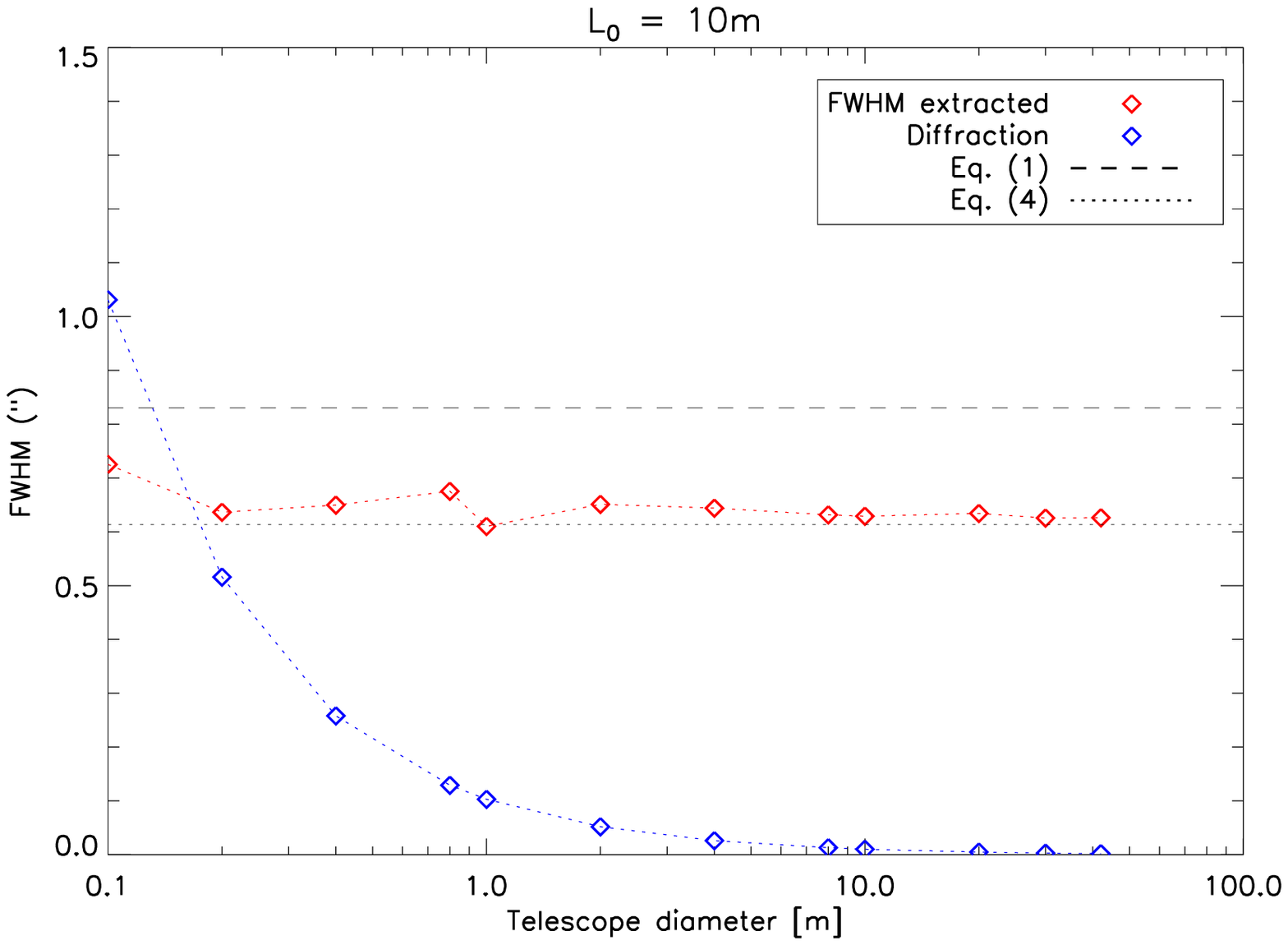}
\includegraphics[width=9cm]{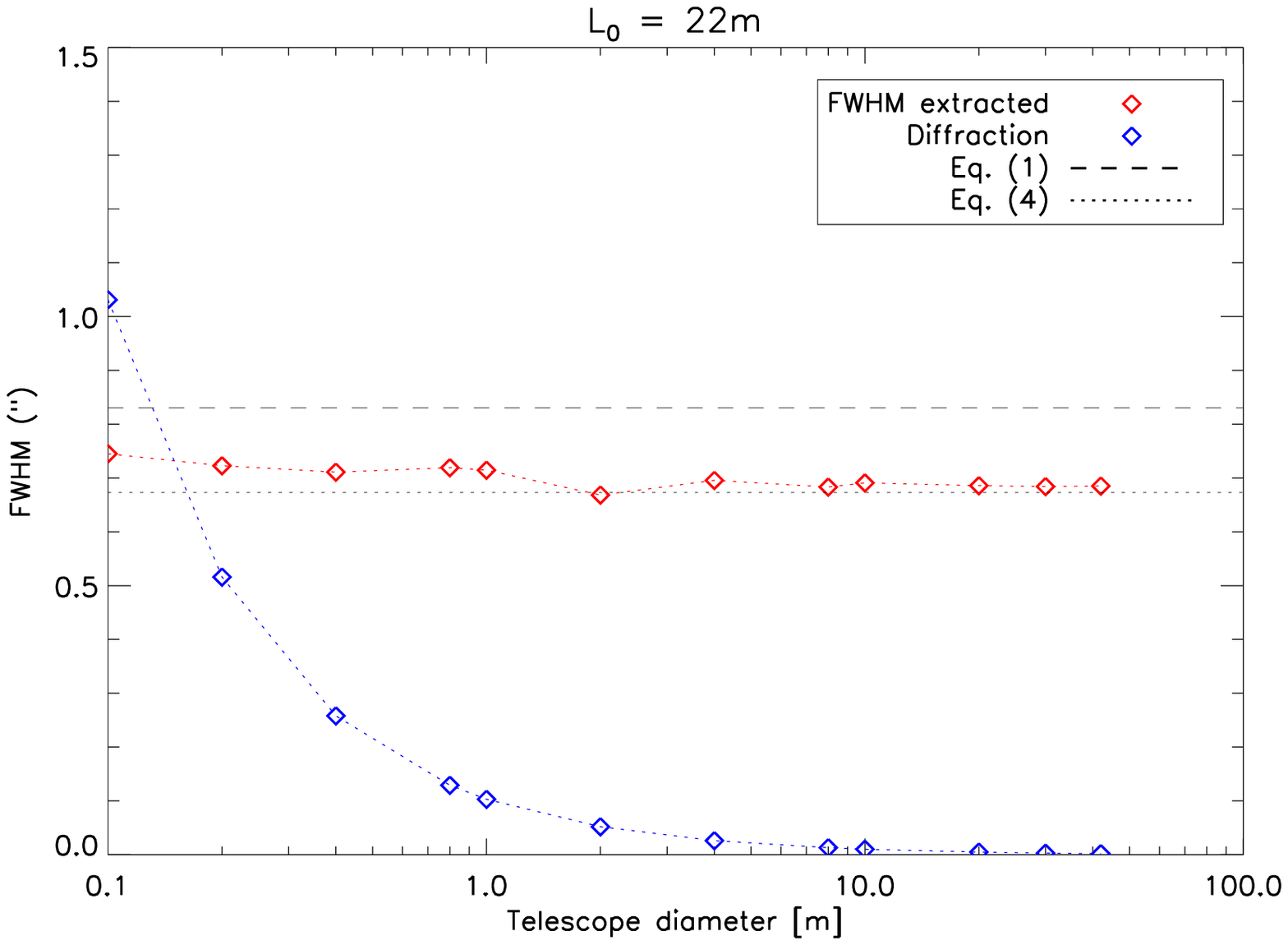}
\includegraphics[width=9cm]{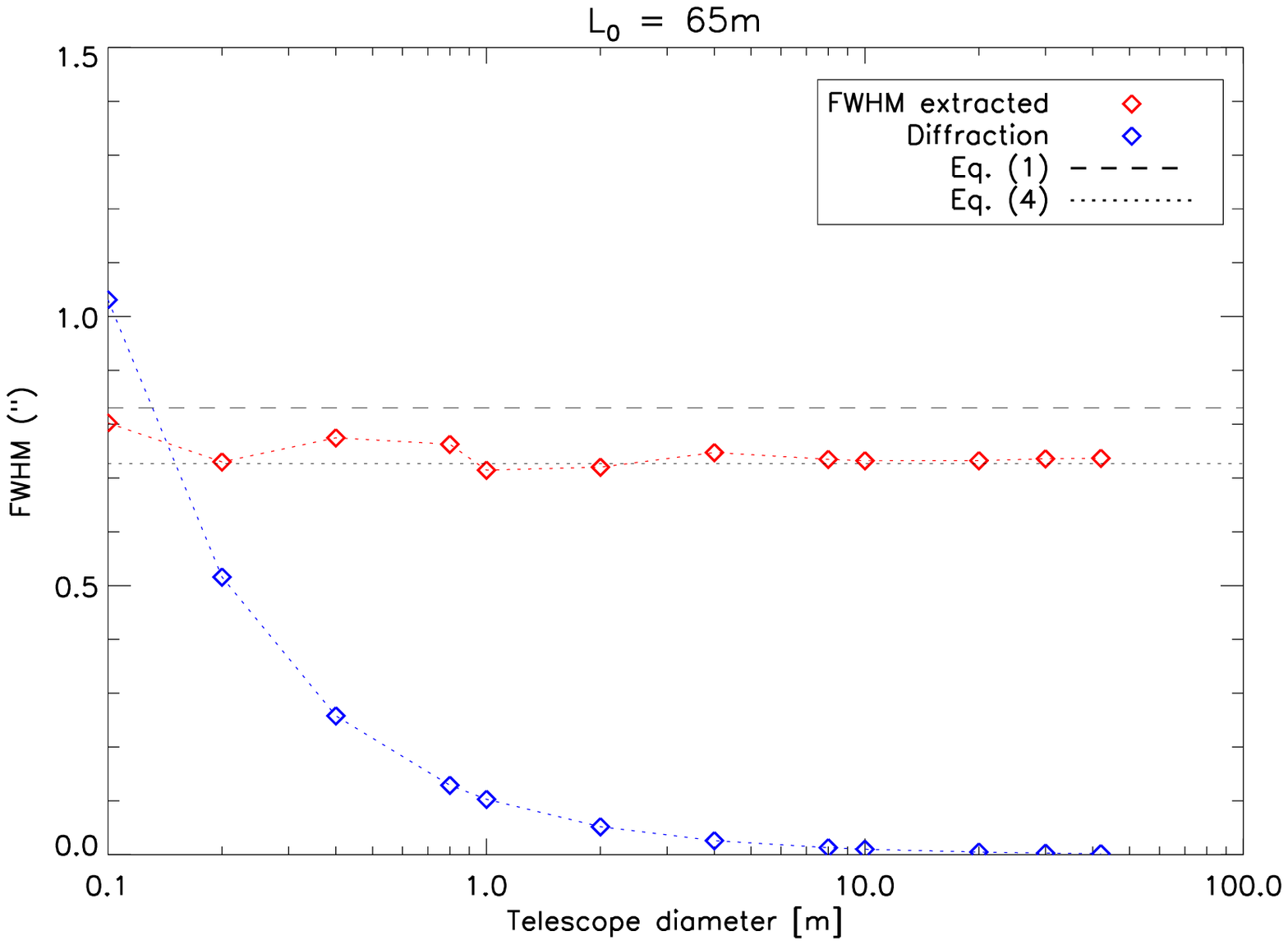}
\includegraphics[width=9cm]{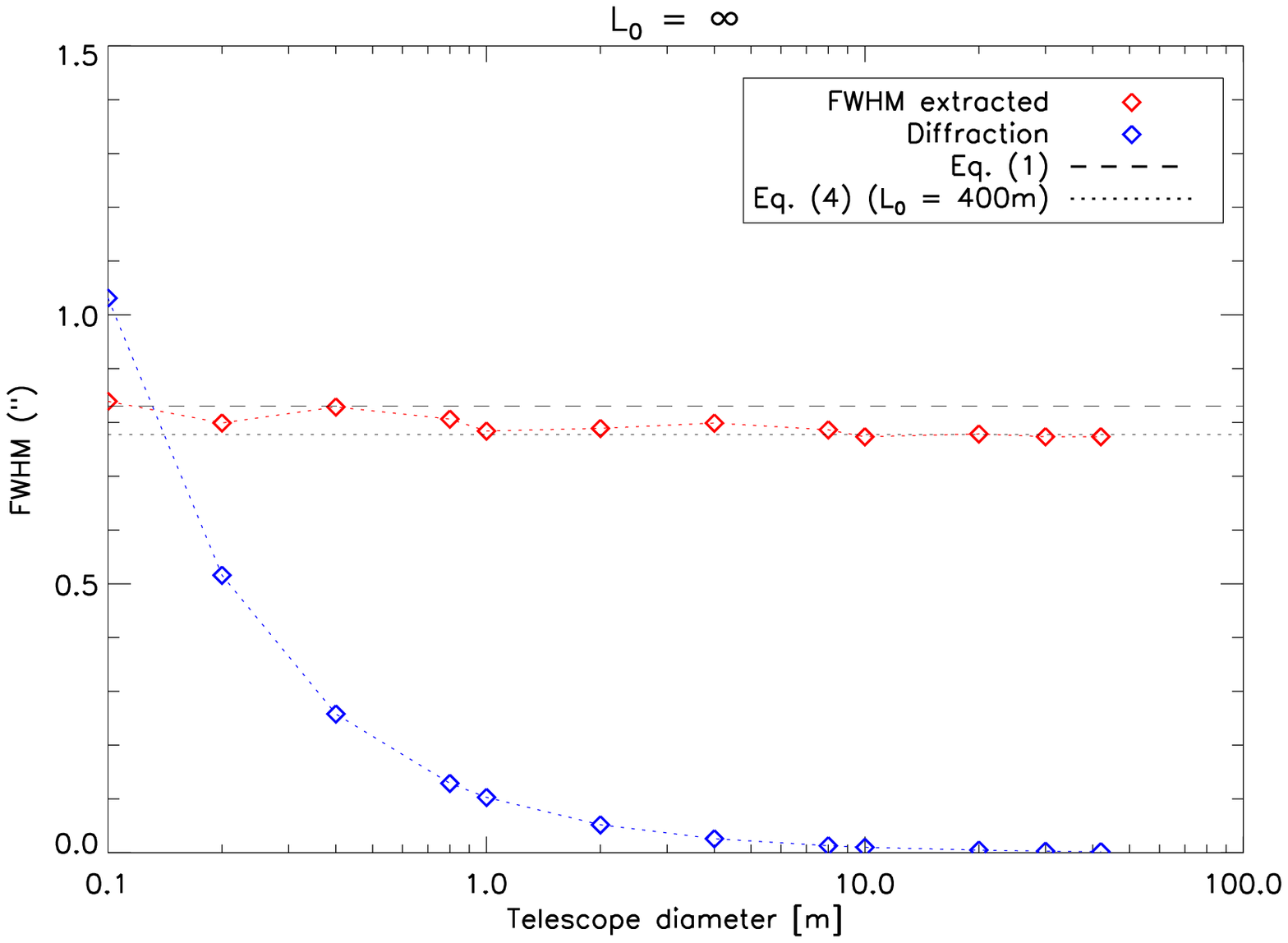}
\end{center}
\caption{The   atmospheric  FWHM   of  simulated   long-exposure  PSFs
  $\theta_{\rm a}$  versus telescope diameter for  several $L_0$ values
  ($\lambda$=$0.5\mu$m,  $\varepsilon_0 =  0.83  \arcsec$).  The  blue
  curves trace the diffraction FWHM $\theta_{\rm dif}= \lambda / D$.}
  \label{Simul012}
\end{figure*} 

Several investigations have  been carried out on the  phase screens to
ascertain  that  their  statistics  correspond  indeed  to  the  input
parameters  $r_0$  and $L_0$.   For  example,  we  compared the  phase
variance, and the variances of the first $100$ Zernike coefficients for
$D = 42\,m$ ($r_{0}$=12.12\,cm  and $L_{0}=65$\,m) with their expected
values given by  \citet{Conan} and found a good  agreement.  The phase
variance matches expectation to  within 1.7$\%$, while the variance of
tip and tilt  components meets their theoretical values to within 1.3 and
0.7$\%$ accuracy, respectively.  The case $L_{0}=\infty$ is particular:
the  variance of  the tip  and tilt  coefficients does  not  fit their
theoretical values,  corresponding instead to a finite  outer scale in
the range  of 200\,m to 500\,m.   This is a consequence  of the finite
size of the simulated phase screens.

Several telescope diameters have been considered ranging from 10\,cm to
42\,m.  The  wavelength domain ranges from the U-band to M-band, while
the seeing ranges from 0.1$\arcsec$ to 1.8$\arcsec$.

All simulations involve Fast  Fourier Transforms (FFT) of $8192 \times
8192$   arrays  to   generate  the   long-exposure  PSFs   (over  1000
realizations).   The  same sets  of  phase  screens  is used  for  all
telescope diameters.  As a result of the very large arrays involved in
order  to handle  both  phase screen  statistics  and aliasing  effect
(e.g. large-aperture cases), small  telescope diameters ($< 1$\,m) may
suffer from coarse pupil sampling.  The effect of speckle structure is
also stronger  for small $D$, causing  a larger random  scatter in the
results.

\subsection{Measurement of the FWHM}

We   determined  the   FWHM  of   the  simulated   long-exposure  PSFs
$\theta_{\rm PSF}$  in  the  following   way.   The  PSFs  were  first
azimuthally  averaged. 
The 10-th order polynomial  was then fitted to this  curve, and the radius
where it crosses the 1/2 of the maximum intensity was determined.  The
outcomes  of  this routine  has  been  compared  to another  algorithm
\citep{kolb}  applied on  the  same set  of  PSFs, and  both gave 
similar values (e.g. $\pm$1 pixel for D = 8m and $L_{0}$ = 22m, i.e. 0.006$\arcsec$).

The  simulated PSFs  are broadened  by  diffraction and  thus are  not
directly comparable  to Eq.~\ref{toko}.  We  approximately account for
this  by subtracting  quadratically the  diffraction  FWHM $\theta_{\rm
  dif} = \lambda / D$,
\begin{equation}
\theta_{\rm a} \approx \sqrt{{\theta_{\rm PSF}}^{2} - {\theta_{\rm dif}}^{2}} .
\end{equation}
This gives a  good approximation to $\varepsilon_{\rm vK}$  as long as
the  diffraction  blur is  small,  $D \gg  r_0$,  but  fails at  small
diameters,  as  mentioned  above,  because the  individual  broadening
factors are not Gaussian. This  explains why our results for small $D$
are inaccurate.

\subsection{Outer scale and telescope diameter}

The first series of simulations  aims at defining the general trend of
atmospheric FWHM  $\theta_{\rm a}$ in large telescopes  in the presence
of finite outer scale. We compare $\theta_{\rm a}$ to Eq. \ref{toko} and  to  the  seeing  $\varepsilon_0$,  fixed  at  $0.83
\arcsec$ in this case.     
Some   results   are   presented in Fig.~\ref{Simul012}.

From Fig.  \ref{Simul012} it is straightforward to see that $\theta_{\rm a} <
\varepsilon_0$ in  all cases,  even for $L_0  = \infty$  because {\em
  all} simulated wave-fronts have finite outer scale. As expected, the
validity of  Eq. \ref{toko} is confirmed, except for
the small $D$ where  our treatment of  diffraction is too crude. All
these  cases correspond $L_0/r_0  > 80$ where the  effect  of the
finite $L_0$ is still mild. 
\begin{figure}[!t]
\begin{center}
\includegraphics[width=9cm]{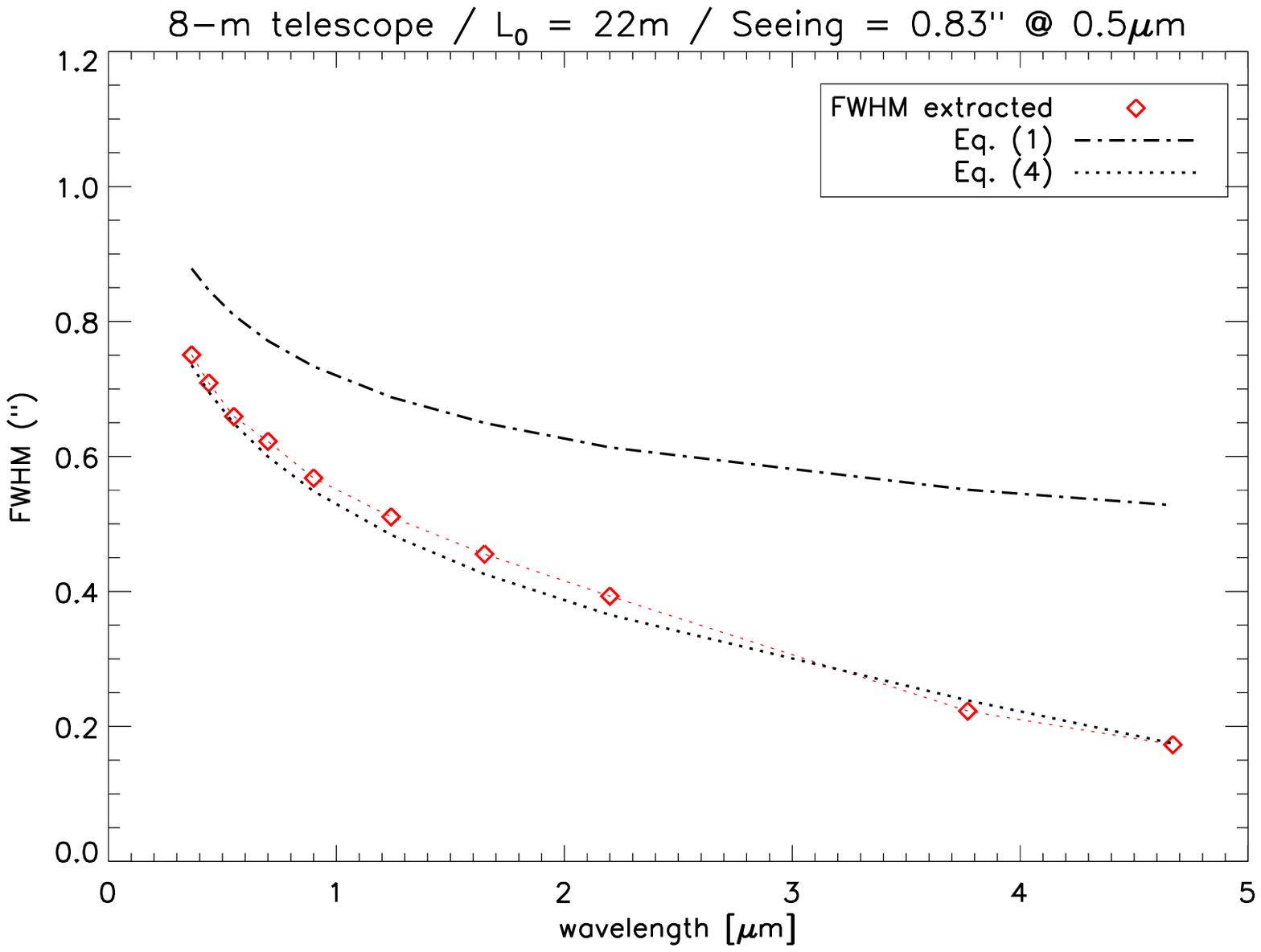}
\includegraphics[width=9cm]{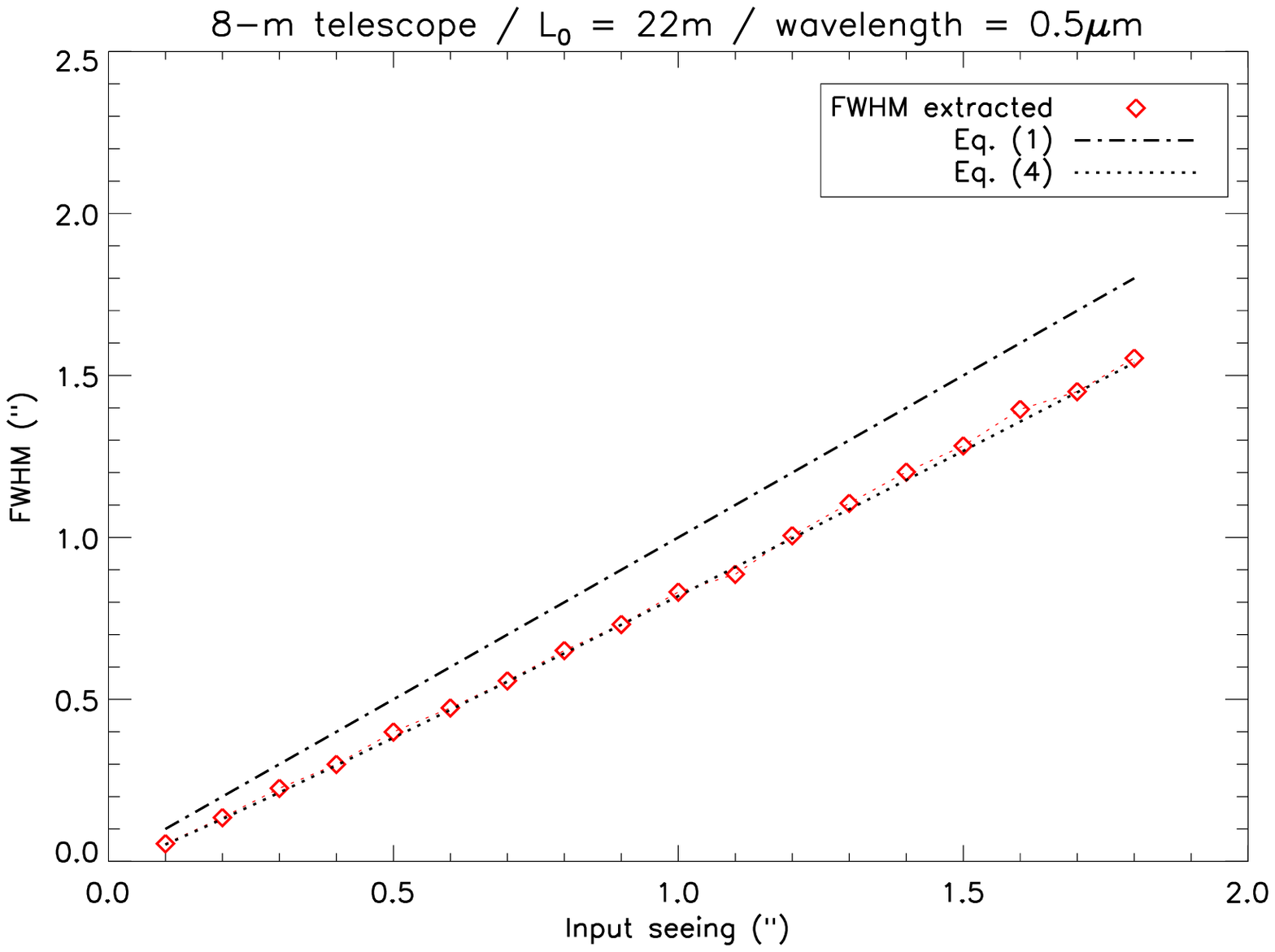}
\end{center}
\caption{Dependence of   $\theta_a$  on wavelength  (top,
  $\varepsilon_0 = 0.83''$) and
  seeing   (bottom,  $\lambda=0.5\mu$m ). Other parameters: $L_0 = 22$\,m,
  $D = 8$\,m. }
\label{Simul03}
\end{figure} 

\subsection{Wavelength and seeing dependence}

For  the  second series  of  simulation,  we  have considered  an  8-m
telescope, a fixed outer scale $L_0 = 22$\,m, and 0.83$\arcsec$ seeing
at 0.5$\mu$m, while  the imaging wavelength is varying from  the U-band to the
M-band  (from 0.365 to  4.67\,$\mu$m). 
The  results are  presented in Fig.~\ref{Simul03}, top.  Note the stronger  dependence of $\theta_{\rm
  a}$  on wavelength,  compared  to the  Kolmogorov  case.  The  third
series  of simulation  considers the  same  $L_0$ and  $D$, while  the
seeing  conditions are  evolving (Fig.   \ref{Simul03},  bottom).  The
agreement  with  Eq. \ref{toko} is  demonstrated  for  both  wavelength
($L_0/r_0 > 10$) and seeing dependence ($L_0/r_0 > 20$).

\subsection{Discussion}

The previous subsections gave general results for the atmospheric FWHM
in the  presence of a finite  outer scale. In order  to relate these
results to actual situation, we discuss the particular case of the 8-m
Very Large Telescope at  Paranal (Chile), and assuming standard seeing
conditions  (0.83$\arcsec$  at 0.5\,$\mu$m),  and  median outer  scale
value  ($L_0=22$\,m,  i.e.   $L_0/r_0$   =  180).   Results  shown  in
Fig.~\ref{Simul03} indicate  that the  FWHM of the  von K\'{a}rm\'{a}n
PSF is reduced by $19\%$ compared to standard theory ($\varepsilon_0$)
in the  visible.  It is even  more dramatic in the  near-IR, where the
FWHM ($\varepsilon_{vK}$)  is reduced  by 29.7$\%$ (H-band)  and 36.3$\%$
(K-band).

\begin{figure}[!t]
\begin{center}
\includegraphics[width=9cm]{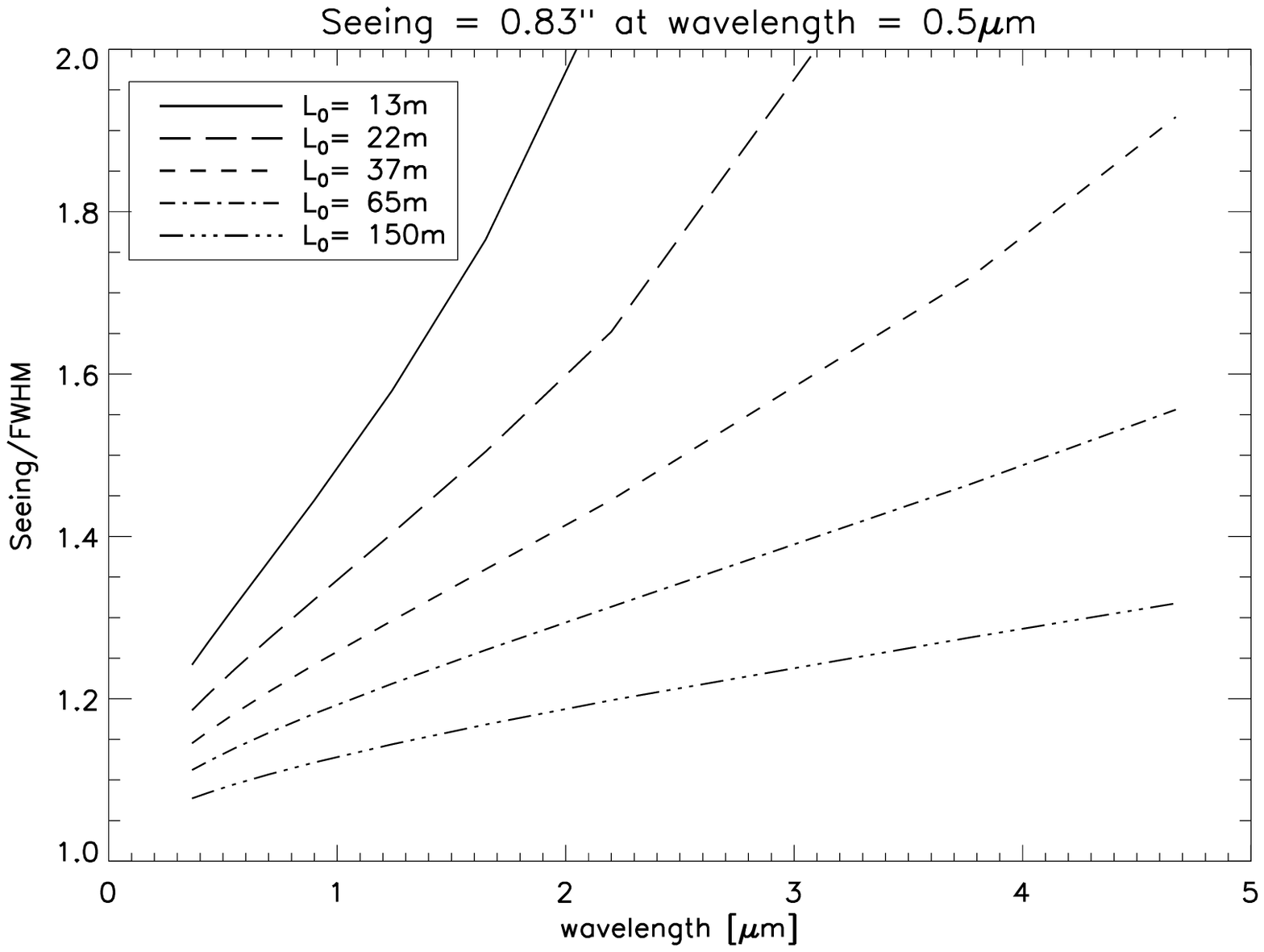}
\includegraphics[width=9cm]{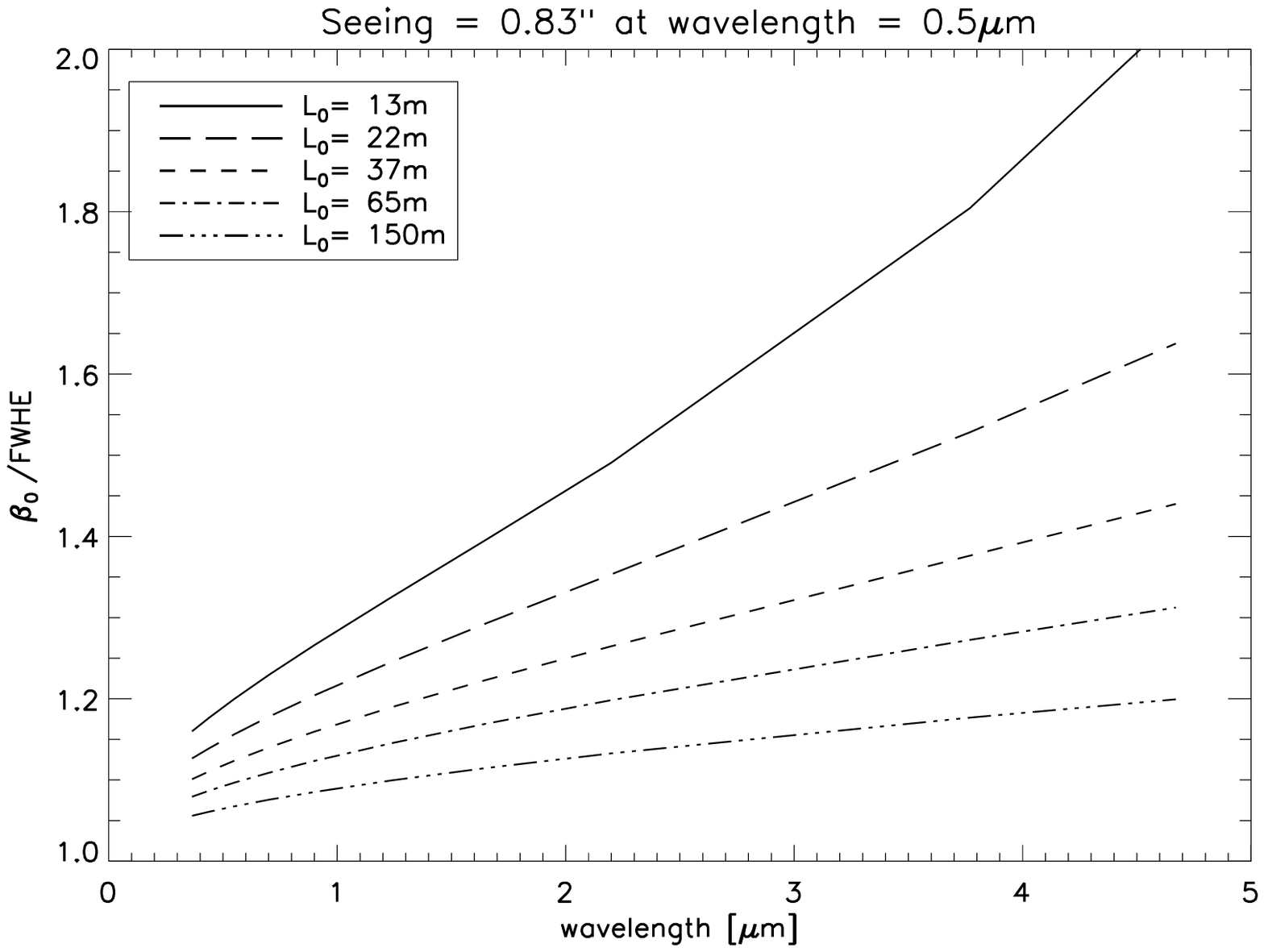}
\end{center}
\caption{Top: Ratio between seeing  $\varepsilon_0$ and  FWHM ($\varepsilon_{vK}$)  as a  function of the  wavelength for  several $L_0$
  cases. Bottom: similarly, ratio between $\beta_0$ and FWHE ($\beta_{vK}$). For both plots, seeing is 0.83$\arcsec$ at 0.5$\mu$m.}
\label{Simul04}
\end{figure}

In   the   same    way,   Fig.~\ref{Simul04} (top)   quantifies   ratio   of
$\varepsilon_0$   by   $\varepsilon_{vK}$   (i.e.   Eq.~\ref{eq1}   to
Eq.~\ref{toko}) but for several  $L_0$ values ranging from the Paranal
median value (including the 1-$\sigma$ outer scale values, 13 and 37\, m) to 150\,m.  The difference with  the standard Kolmogorov
seeing formula  is substantial and can  exceed a factor of  two in the IR.
Likewise, Fig.~\ref{Simul04} (bottom) compares ratio of $\beta_0$   by   $\beta_{vK}$.

\section{Resolution under partial compensation}
\label{sec:AO}
In analogy with the finite outer scale impact, we discuss here the consequences of a reduction of the low-frequency content of the phase perturbation spectrum generated by AO partial correction, or tip-tilt compensation.

The  purpose  of  Adaptive   Optics  (AO)  systems  is  to  compensate
atmospheric  wave-front distortions  and to  reach diffraction-limited
resolution. To do this, the actuator spacing $d$ (or an equivalent measure
of AO compensation  order) must be of the order of  $2 r_0$ or smaller
\citep{Roddier98}.   However, AO  systems  which  do  not fulfill  this
condition still improve  the resolution. A good example  is the use of
the low-order  AO system PUEO for observations  at visible wavelengths
\citep{PUEO}. A  resolution gain up to  two times has been  reported. To our
knowledge, shrinking of the atmospheric PSF under partial compensation
has not been explored in a systematic way. 

Residual  wave-fronts  after  AO compensation  contain  high-frequency
ripple,  whereas the  perturbations at  spatial frequencies  less than
$f_c  = 1/(2  d)$ are  corrected.  This  can be  modeled  by high-pass
filtering of  the atmospheric  PSD.  The form  of this  filter varies,
depending on the AO system.  The calculations here only illustrate the
principle and should be repeated for each AO system if an exact result
is sought.   We model the  AO compensation by a  multiplicative factor
$F$:
\begin{equation}
F(x ) = x/(1+x), \;\;\; x=(|{\bf f}| /f_c)^m
\label{eq:F}
\end{equation}
with $m  = 6$. The PSD (Eq. \ref{eq:L0})  is multiplied by $F$,  the SF is
calculated and used to compute the residual PSF in the same way as for
the vK spectrum. The SF saturates  at $r \gg d$, reaching the value $2
\sigma_{\phi}^2 =  0.62 (d/r_0)^{5/3}$.  The  shape of the SF  and the
saturation  value depend on  the filter  $F(x)$. We  experimented with
several filters  and have chosen  Eq. \ref{eq:F} with $m=6$  because it
  matches  approximately the  known formula  $ \sigma_{\phi}^2  = 0.35
  (d/r_0)^{5/3}$, \citet{Roddier98}.  
Comparing  this to the saturation
  level of  the vK SF, $0.17  (L_0/r_0)^{5/3}$, we may  state that the
  effective outer scale of the residual wavefront is $\sim 2 d$.

Once  the SF  saturates at  $2 \sigma_{\phi}^2$,  the  atmospheric OTF
reaches a  constant level $ T_{\rm  min} = \exp (-  \sigma^2)$. We can
represent such an  OTF as a sum of the constant  term and a decreasing
part. This corresponds to the  sum of a diffraction-limited PSF scaled
by $S = \exp (- \sigma^2)$ and  a wide residual halo. The shape of the
halo can therefore be calculated by replacing the atmospheric OTF with
$(T_a - T_{\rm min})/ (1 - T_{\rm min})$ out to the distance where the
minimum $T_{\rm  min}$ is reached, and  setting it to  zero for larger
frequencies.  Note that we  re-normalize the  halo OTF  to one  at the
coordinate origin.

\begin{figure}[!t]
\begin{center}
\includegraphics[width=9cm]{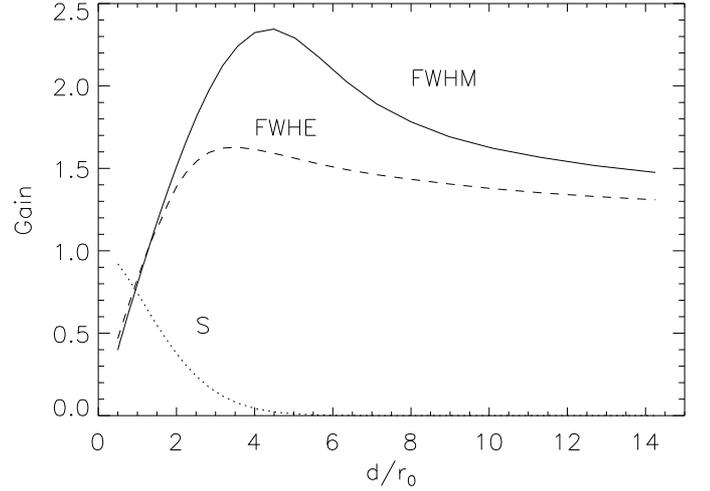}
\end{center}
\caption{Gain in the FWHM (full  line) and FWHE (dotted line) diameter
  of  the PSF  halo (compared  to the  Kolmogorov PSF)  resulting from
  partial AO  compensation (outer scale not included). The dotted line shows  the coherent energy
  $S$.  }
\label{fig:AO}
\end{figure}

The FWHM $\theta_{\rm AO}$ and the diameter of a circle containing half
the  energy  (FWHE) have  been  computed  for  the halo  of  partially
compensated PSFs.  We compare  these parameters to the non-compensated
(Kolmogorov) PSFs in Fig.~\ref{fig:AO}. Even when the actuator spacing
$d$ is much larger than $r_0$  and the AO system does not perform well
in the classical sense ($S \approx 0$), the gain in FWHM and FWHE is
already  substantial. Maximum  resolution gain  $\varepsilon_0  / \theta_{\rm
  AO}$ is reached  at $d/r_0 \sim 4$, when the  coherent PSF core is still
very weak.   As the  compensation order increases  further, increasing
fraction  of energy goes  into the  core, the PSF  halo becomes
weaker and  wider. At  small $d$  and high $S$,  the halo  becomes even
wider  than  the un-compensated  atmospheric  PSF,  being produced  by
residual phase errors at spatial scales smaller than $r_0$.

Tip-tilt correction is a particular case of low-order AO compensation.
It is  well known that maximum  resolution gain is  achieved at $D/r_0
\sim 3.6$ \citep{Fried66}. The gain  studied here does not depend on the
telescope  diameter $D$,  but  rather on  the dimensionless  parameter
$d/r_0$, in full analogy with the effect of the outer scale. 

All three effects  -- outer scale, partial AO  correction and tip-tilt
compensation  --  reduce  the   low-frequency  content  of  the  phase
perturbations  spectrum.    When  they  act  together,   the  gain  in
resolution over Kolmogorov turbulence  is not cumulative. For example,
with finite  outer scale the tip-tilt fluctuations  become smaller and
their  correction  achieves a smaller  resolution  gain. Similarly,  the
resolution gain from partial AO correction (Fig.~\ref{fig:AO}) in fact
will be less because of the finite $L_0$.

\section{Conclusions and discussion}
\label{conclusion}

This study is largely motivated by the confusion between seeing and
the FWHM of long-exposure images in large telescopes, also often
called {\em delivered image quality} (DIQ). In an ideal large telescope
(no aberrations, internal turbulence and wind shake), the DIQ is
always less than  predicted by the standard theory, owing
to the finite turbulence outer scale. 

The  seeing  is usually  measured  by  the  Differential Image  Motion
Monitors   \citep[][DIMMs]{Martin87,  Sarazin90}.    This   method  is
sensitive to small-scale wave-front distortions and provides estimates
of     $r_0$    which    are     almost    independent     of    $L_0$
\citep{Ziad94}  \footnote{Seeing  monitors based  on  the absolute
  image motion  are affected by  finite $L_0$, giving  biased, larger
  $r_0$ values.}.   Using the  standard theory, we  will over-estimate
the FWHM  expected in a  large telescope. As  the PSF is  broadened by
non-atmospheric  factors, this  mismatch can  hide  telescope defects.
This is  particularly dangerous in  the IR, where the  difference with
the   standard  theory   is  large.    Therefore,  reaching   a  truly
seeing-limited  telescope  performance in  the  IR requires  mandatory
account of finite  $L_0$. Stated in other words,  our telescopes could
perform better than we think  they should based on the standard theory
and DIMM measurements.

If, on the  other hand, we want to deduce  atmospheric seeing from the
width of the long-exposure PSF,  the situation is reversed. The actual
seeing is  worse than we think it  is.  
The effect of  finite $L_0$ is apparent for {\em all} telescope diameters. 
A simultaneous measurement of $L_0$ is thus required to be accurate.
Estimating seeing from the width  of  the  spots  in active-optics  Shack-Hartmann  sensor  (long
exposures) should be done with these circumstances in mind. 


As  internal  telescope  defects  and  outer  scale  act  in  opposite
directions,  they can  partially compensate  each other.  An agreement
between DIMM  measurements and DIQ can  thus be found  where it should
not be \citep{Sarazin90}.
By comparing  simultaneous PSFs  at visible and  in the mid-IR,  it is
possible  to  extract   two  parameters,  $\varepsilon_0$  and  $L_0$,
assuming that  the telescope's  contribution to the  image degradation
can be neglected \citep{Toko2007}.


\bibliography{MyBiblio3}
\end{document}